\pdfoutput=1 % arXiv admin added this line
\documentclass[apjl]{emulateapj}

\usepackage{epstopdf}

\newcommand{\deluxetablestar}{deluxetable*} % for emulateapj
\newcommand{\vlsr}{\ensuremath{v_\mathrm{LSR}}}
\newcommand{\K}{\ensuremath{ \, \mathrm{K}}}
\newcommand{\pc}{\ensuremath{ \, \mathrm{pc}}}
\newcommand{\kpc}{\ensuremath{\, \mathrm{kpc}}}
\newcommand{\cm}{\ensuremath{ \, \mathrm{cm}}}
\newcommand{\cucm}{\ensuremath{ \cm^{-3}}}
\newcommand{\cmsix}{\ensuremath{ \, \mathrm{cm}^{-6}}}
\newcommand{\s}{\ensuremath{ \, \mathrm{s}}}
\newcommand{\kms}{\ensuremath{ \, \mathrm{km} \, \s^{-1}}}

\newcommand{\R}{\ensuremath{ \, \mathrm{R}}}
\newcommand{\sr}{\ensuremath{ \, \mathrm{sr}}}

\newcommand{\Hneutral}{\ensuremath{\textrm{\ion{H}{1}}}}
\newcommand{\Hplus}{\ensuremath{\mathrm{H}^+}}
\newcommand{\Lneutral}{\ensuremath{L_{\Hneutral}}}
\newcommand{\Lionized}{\ensuremath{L_{\Hplus}}}
\newcommand{\NH}{\ensuremath{N_{\Hneutral}}}
\newcommand{\nneutral}{\ensuremath{n_0}}

\newcommand{\ha}{\ensuremath{\mathrm{H} \alpha}}
\newcommand{\hb}{\ensuremath{\mathrm{H} \beta}}
\newcommand{\nii}{\ensuremath{\textrm{[\ion{N}{2}]}}}
\newcommand{\sii}{\ensuremath{\textrm{[\ion{S}{2}]}}}
\newcommand{\oiii}{\ensuremath{\textrm{[\ion{O}{3}]}}}
\newcommand{\EM}{\ensuremath{\textrm{EM}}}

\newcommand{\N}{\ensuremath{\mathrm{N}}}
\renewcommand{\H}{\ensuremath{\mathrm{H}}}
\newcommand{\Sulfur}{\ensuremath{\mathrm{S}}}

\newcommand{\mc}{\multicolumn{2}{c}}
\newcommand{\tnma}{\tablenotemark{a}}
\newcommand{\tnmb}{\tablenotemark{b}}
\newcommand{\tnmc}{\tablenotemark{c}}
\newcommand{\tnmd}{\tablenotemark{d}}

\begin{document}

\author{Alex S. Hill, L. Matthew Haffner, and Ronald J. Reynolds}
\affil{Department of Astronomy, University of Wisconsin-Madison, Madison, WI 53706}
\email{hill@astro.wisc.edu, haffner@astro.wisc.edu, reynolds@astro.wisc.edu}
\title{Ionized gas in the Smith Cloud}
\submitted{ApJ, in press}

\begin{abstract}
We present Wisconsin \ha\ Mapper observations of ionized gas in the Smith Cloud, a high velocity cloud which Lockman et al.\ have recently suggested is interacting with the Galactic disk. Our \ha\ map shows the brightest \ha\ emission, $0.43 \pm 0.04 \R$, coincident with the brightest \ion{H}{1}, while slightly fainter \ha\ emission ($0.25 \pm 0.02 \R$) is observed in a region with \ion{H}{1} intensities $< 0.1$ times as bright as the brightest \ion{H}{1}. We derive an ionized mass of $\gtrsim 3 \times 10^6 M_\odot$, comparable to the \ion{H}{1} mass, with the \Hplus\ mass spread over a considerably larger area than the \Hneutral. An estimated Galactic extinction correction could adjust these values upwards by $40 \%$. \ha\ and \sii\ line widths towards the region of brightest emission constrain the electron temperature of the gas to be between $8000 \K$ and $23000 \K$. A detection of \nii\ $\lambda 6583$ in the same direction with a line ratio $\nii / \ha = 0.32 \pm 0.05$ constrains the metallicity of the cloud: for typical photoionization temperatures of $8000-12000 \K$, the nitrogen abundance is $0.15-0.44$ times solar. These results lend further support to the claim that the Smith Cloud is new material accreting onto the Galaxy.
\end{abstract}


\maketitle

\section{Introduction}

The infall of gas is a significant driver of the evolution of galaxies. Without infall, there is insufficient gas in the interstellar medium of the Galaxy to sustain the observed star formation rate for more than $\sim 1 \textrm{ Gyr}$. Also, the narrow distribution of metallicities of long-lived stars in the solar neighborhood implies that the metallicity of the ISM is roughly constant over time \citep[the so-called ``G-dwarf problem'';][]{v62}. In the Milky Way, high velocity clouds (HVCs) provide direct observational evidence for infalling low metallicity gas \citep{wv97,wyw08}.

The Smith Cloud is an excellent example of the active infall of material. Also called the Galactic center positive (GCP) complex, the Smith Cloud is an HVC with a radial velocity near $+100 \kms$ with respect to the local standard of rest \citep{s63}. \citet[hereafter L08]{lbh08} presented a survey of \ion{H}{1} 21~cm emission from the cloud using the Green Bank Telescope, and \citet{bvc98} and \citet[hereafter P03]{pbv03} identified weak \ha\ and \nii\ emission associated with the cloud.\footnote{Throughout this paper, we use ``\ion{H}{1}'' to refer to both neutral, atomic hydrogen gas and the 21~cm emission from the gas. We use, e.\ g., ``\Hplus'' and ``\N$^+$'' to refer to other ions (ionized hydrogen and singly ionized nitrogen, respectively), with ``\ha'' referring to the Balmer-$\alpha$ emission from recombining ionized hydrogen, \Hplus. Other ionic species notations refer to the emission lines from the gas (e.\ g. ``\nii'' refers to forbidden emission from $\N^+$ gas). The \ion{H}{1} 21~cm intensity is proportional to the \ion{H}{1} column density, whereas the \ha\ intensity is proportional to the line integral of the square of the \Hplus\ density.}
 Like many HVCs, the cloud has a cometary morphology with a bright tip and a diffuse, trailing tail. L08 showed that the cloud is interacting with the Galactic disk and used this information to obtain a kinematic distance of $11.1-13.7 \kpc$, consistent with a stellar absorption distance constraint \citep{wyw08}. Therefore, the position and trajectory of the cloud are both well known: the head is moving towards the plane at $73 \pm 26 \kms$, and the portion of the cloud mapped by L08 has an \ion{H}{1} mass of $\gtrsim 10^6 M_\odot$.

In this paper, we present Wisconsin \ha\ Mapper (WHAM) observations of ionized gas in the cloud, including the first spectroscopic \ha\ map and pointed, velocity-resolved, high signal-to-noise-ratio spectra of the \ha, \nii\ $\lambda 6583$, \sii\ $\lambda 6716$, and \oiii\ $\lambda 5007$ lines.

\section{Observations}

WHAM is a dual-etalon Fabry-Perot spectrometer coupled to a siderostat, designed to observe very faint optical emission lines from diffuse gas. The instrument \citep[described by][]{hrt03} integrates all emission within its $1 \arcdeg$ field of view, sacrificing spatial information to obtain spectra with a resolution of $12 \kms$ over a $200 \kms$-wide window with a sensitivity below $0.1 \R$ in a $30 \s$ exposure.\footnote{$1 \textrm{ Rayleigh} = 10^6 / 4 \pi \textrm{ photons} \cm^{-2} \sr^{-1} \s^{-1}$.} We calibrated the geocentric velocity of the \ha\ emission with a fit of the geocoronal \ha\ line; this calibration is accurate to $\approx 1 \kms$. Other emission lines are calibrated based on the instrument configuration and are accurate to within a few \kms\ \citep{m04}. The observations presented here were obtained with WHAM while it was located on Kitt Peak in Arizona.

\subsection{Pointed observations}

\begin{figure}
%\epsscale{0.8}
\plotone{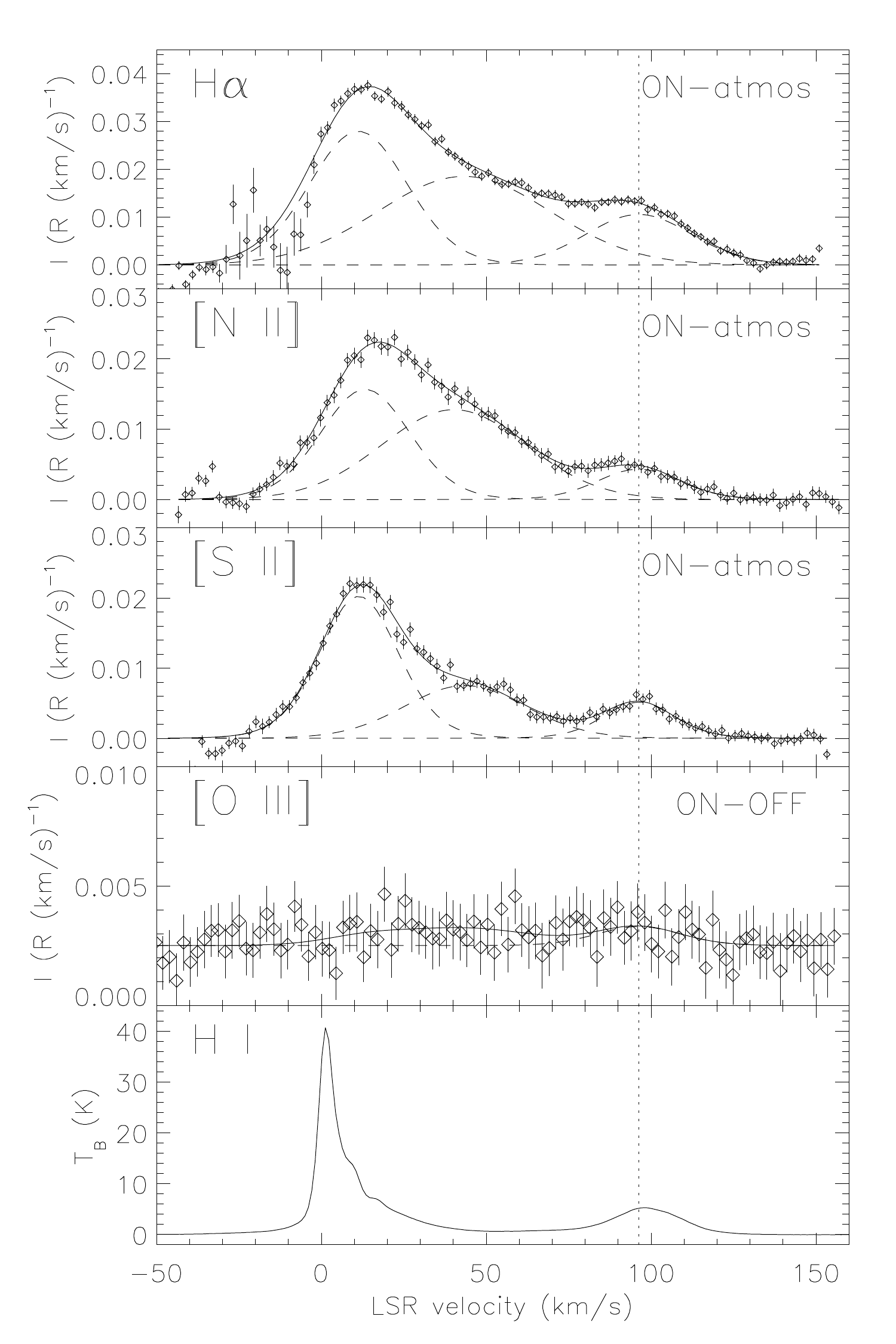}
\caption{Deep spectra of the the Smith Cloud tip (the blue $\times$ in Fig.~\ref{fig:smith_map}). Data are shown as diamonds with error bars; the large uncertainties in the \ha\ spectrum near $\vlsr = -20 \kms$ are due to the subtraction of the relatively bright geocoronal emission. A multiple-component Gaussian fit to the data is shown as a solid line with each component plotted as a dashed line; fit parameters are listed in Table~\ref{tbl:fits}. The average of L08 \ion{H}{1} 21~cm spectra over the WHAM beam is shown in the bottom panel.}
\label{fig:spectra}
\end{figure}

In 2007 October, we obtained spectra towards the tip of the cloud using the following sequence. 120~s ``ON'' observations of the tip ($l=38.6 \arcdeg$, $b=-13.1 \arcdeg$) were alternated with 120~s ``OFF'' observations (chosen based on the \citealt{bvc98} \ion{H}{1} map) for a total of 360~s on source for \ha\ and 480~s on source each for \nii, \sii, and \oiii\ and equal times off source. WHAM spectra have a substantial contribution from the bright geocoronal \ha\ line and numerous fainter, mostly unidentified atmospheric emission lines. \citet{hrh02} constructed an empirical template of the faint lines near \ha, which are stable relative to each other but vary in intensity, partly dependent upon zenith angle. Using this technique for other lines, we fit the OFF spectra of each line to a single atmospheric template, which we then subtracted from the ON spectra to obtain spectra consisting purely of astronomical emission. Summed, atmosphere-subtracted ON spectra for each line are shown in Figure~\ref{fig:spectra}.

\begin{deluxetable}{lll}
\tablecolumns{3}
\tablecaption{Gaussian fit parameters of deep spectra}
\tablewidth{0pt}
\tablehead{Mean (km s$^{-1}$) & FWHM (km s$^{-1}$) & $I$ (R)}
\startdata
\cutinhead{\ha, Smith mean fixed to \sii. $\chi_r^2 = 1.2$.}
$+11.2 \pm  0.5$ & $30.3 \pm  2.3$ & $1.06 \pm 0.20$ \\
$+43.2 \pm  4.5$ & $58.3 \pm  7.5$ & $1.21 \pm 0.22$ \\
$+96.2$\tnma     & $33.4 \pm  2.4$ & $0.43 \pm 0.04$ \\
\cutinhead{\nii, Smith mean fixed to \sii. $\chi_r^2 = 0.5$.}
$+13.3 \pm  1.0$ & $26.9 \pm  3.2$ & $0.54 \pm 0.17$ \\
$+40.2 \pm  5.7$ & $48.2 \pm  7.5$ & $0.71 \pm 0.18$ \\
$+96.2$\tnma     & $24.4 \pm  3.7$ & $0.14 \pm 0.02$ \\
\cutinhead{\sii\ 3-component free fit. $\chi_r^2 = 0.6$.}
$+11.3 \pm  0.8$ & $22.9 \pm  1.6$ & $0.62 \pm 0.07$ \\
$+43.5 \pm  4.0$ & $39.1 \pm  8.2$ & $0.35 \pm 0.07$ \\
$+96.2 \pm  1.1$ & $19.7 \pm  3.3$ & $0.14 \pm 0.02$ \\
\cutinhead{\oiii, Smith mean fixed to \sii. $\chi_r^2 = 0.3$.}
$+13.3$\tnma     & $25.0$\tnma     & \nodata\tnmb \\
$+43.5$\tnma     & $45.0$\tnma     & \nodata\tnmb \\
$+96.2$\tnma     & $25.0$\tnma     & $0.026 \pm 0.014$ \\
\cutinhead{Adopted values for the Smith Cloud tip}
\ha  & $33.4 \pm 2.4$ & $0.43 \pm 0.04$\tnmc \\
\nii & $24.4 \pm 3.7$ & $0.14 \pm 0.02$\tnmc \\
\sii & $19.7 \pm 3.3$ & $0.14 \pm 0.02$\tnmc \\
\oiii & $25$\tnma     & $<0.07 \, (3 \sigma)$ 
\enddata
\tablenotetext{a}{Values with no uncertainty were fixed in the fit.}
\tablenotetext{b}{\oiii\ components at $\vlsr < +50 \kms$ account for excess residual emission in the ON$-$OFF spectrum and have no physical significance.}
\tablenotetext{c}{Intensities are not corrected for extinction, as discussed in \S~\ref{sec:extinction}.}
\label{tbl:fits}
\end{deluxetable}

We fit the atmosphere-subtracted spectra with three Gaussians convolved with the instrument profile: one for local gas near $\vlsr = +10 \kms$, one near $\vlsr = +40 \kms$, corresponding to the Sagittarius Arm of the Galaxy, and one for the Smith Cloud near $+100 \kms$. Solutions are listed in Table~\ref{tbl:fits}, with our adopted line widths and intensities listed in the bottom of the table. We first performed an unconstrained fit. Blending of the emission from the Smith Cloud and the relatively wide Sagittarius Arm component is the dominant source of uncertainty in constraining the multiple-Gaussian fits, although the width and intensity of the Smith Cloud component were well-defined in all three lines. The relatively narrow \sii\ line provides the best constraint on the mean velocity of each component, so we fixed the Smith Cloud velocity in the \ha\ and \nii\ spectra to the $+96.2 \kms$ of the \sii\ line and performed a second fit. The uncertainties are statistical and do not account for systematic effects; for example, any unidentified components in the spectrum would cause the widths in our fit to be overestimated.

For the \oiii\ spectrum, we have no suitable atmospheric template for the weak emission at the Smith Cloud velocities. Figure~\ref{fig:spectra} shows an ON$-$OFF spectrum for this line; no atmospheric template is subtracted. We fit the spectrum by fixing the velocities of all three components to those observed for \sii\ and the widths at values similar to the observed width of the \nii\ components. This yields a best fit \oiii\ intensity in the Smith Cloud tip of $0.026 \pm 0.014 \R$, with the reduced $\chi^2$ parameter $\chi_r^2 = 0.34$. However, fitting a flat line to the spectrum (no \oiii\ emission) yields nearly as good a fit, with $\chi_r^2 = 0.40$. This result is a $3 \sigma$ upper limit of $0.07 \R$, fainter than a previous \oiii\ upper limit of $0.12 \R$ \citep{bvc98}.

\subsection{Maps} \label{sec:maps}

\begin{figure*}
%\epsscale{1.0}
\plotone{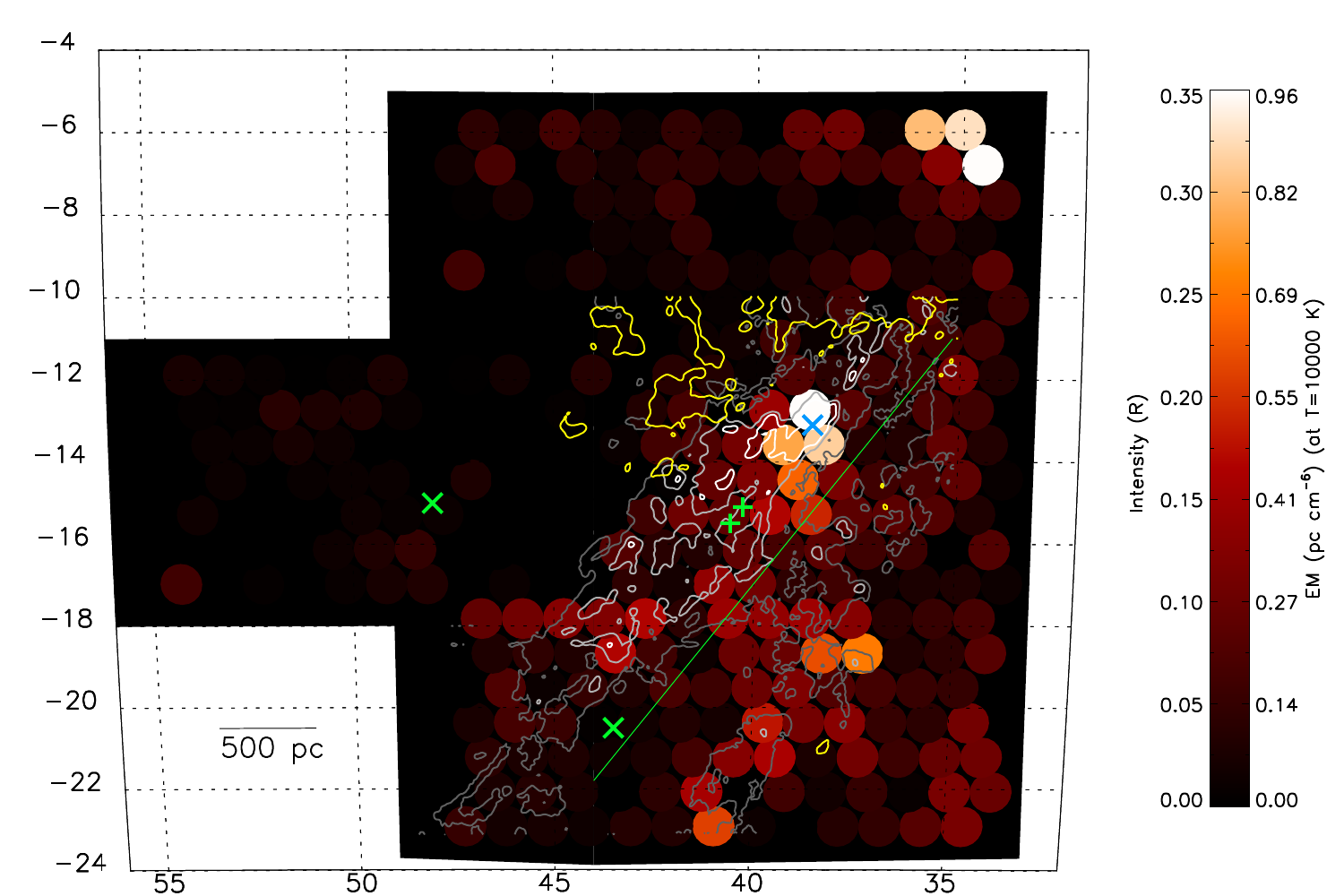}
\caption{Map of the Smith Cloud in \ha\ in Galactic coordinates. Intensities are derived from the area of a Gaussian fitting the \ha\ spectrum centered at $+96.2 \kms$ with a width of $30 \kms$. Greyscale contours of \ion{H}{1} column density in the L08 data, integrated from $+70$ to $+145 \kms$, are shown at $2$ (dark grey), $10$ (light grey), and $20$ (white) $\times 10^{19} \cm^{-2}$. Yellow contours show intermediate velocity \ion{H}{1} integrated from $+50$ to $+60$ \kms\ at $4 \times 10^{19} \cm^{-2}$. A blue $\times$ sign indicates the postion of the on-source deep exposures, green $\times$ signs show the off-source positions, and green $+$ signs show the positions of the P03 spectra. The bright \ha\ emission near $(35 \arcdeg, -6 \arcdeg)$ is the edge of a low-extinction window into the inner Galaxy, discussed by \citet{mr05}. The scale bar assumes $D=12.4 \kpc$. The green line is discussed in \S~\ref{sec:discussion}.}
\label{fig:smith_map}
\end{figure*}

\begin{figure}
\plotone{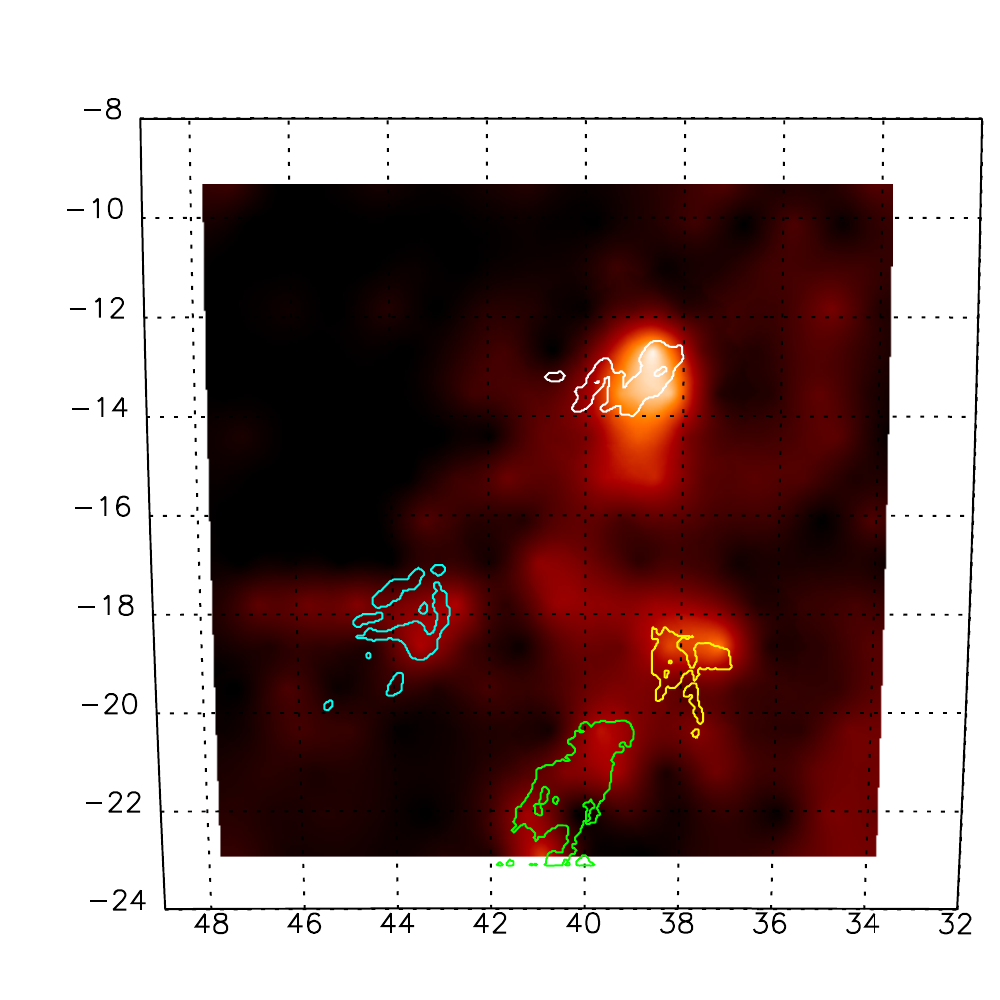}
\caption{Smoothed map of the Smith Cloud in \ha, a portion of the area shown in Fig.~\ref{fig:smith_map}. Contours identify the chosen \ion{H}{1} boundaries of the clouds identified in Table~\ref{tbl:mass}.}
\label{fig:smith_clumps}
\end{figure}

In 2007 May and June, we obtained a map of the \ha\ emission from the Smith Cloud region using WHAM with the ``block'' mapping technique \citep{hrt03}. Observations of each sightline within a $7 \arcdeg \times 7 \arcdeg$ block were obtained in succession with $60 \s$ exposures for each pointing. One-degree beams were observed on a grid with $\Delta l = 0.98\arcdeg / \cos b$ and $\Delta b = 0.85 \arcdeg$, undersampling the image. We used an average spectrum of all pointings within the block to characterize the level of the atmospheric template, which we then subtracted from each spectrum. We fit each atmosphere-subtracted spectrum with the sum of (1) a single Gaussian to account for the geocoronal \ha\ and (2) three Gaussians near LSR velocities of $+10$, $+40$, and $+96.2 \kms$, as observed in the deeper, pointed observations. The widths of all three components were fixed to $30 \kms$, typical of interstellar \ha\ emission and similar to the derived line widths in the pointed observation.

In each fit, the means of the lower-velocity components were allowed to vary freely. Not all beams in the map contain emission at the velocity of the Smith Cloud. We tried fitting the full profile allowing the mean of the $+96.2 \kms$ component to vary, but in sightlines with low or zero emission from the Smith Cloud, the lowest $\chi^2$ fit occurred with that component fitting a residual from lower-velocity Galactic emission instead. To recover the most accurate intensity distribuion of the ionized component of the Smith Cloud, we fixed this component to a mean of $+96.2 \kms$ to trace even the faintest \ha\ emission associated with the cloud near this velocity. The narrower \sii\ line is better suited to a full exploration of the velocity field in the Smith Cloud, but our current data set only has \sii\ data along one sightline.

\subsection{Extinction} \label{sec:extinction}

We estimate the extinction towards the cloud using the mean \ion{H}{1} column density in the direction of the WHAM observations integrated over $-20$ to $+50 \kms$, $\langle \NH \rangle = 9 \times 10^{20} \cm^{-2}$; based on its velocity, this emission should be predominantly foreground. Assuming the mean ratio of color excess to \ion{H}{1} column density found by \citet{bsd78}, we estimate $E(B-V) = 0.15 \textrm{ mag}$. If the dust follows the extinction curves determined by \citet{ccm89} with the standard diffuse ISM value of $R_V \equiv A(V)/E(B-V) = 3.1$, this yields a total extinction at \ha\ of $A(\ha) = 0.38 \textrm{ mag}$. The scatter in the foreground \ion{H}{1} column density leads to a standard deviation of $\sigma_{A(\ha)} = 0.11 \textrm{ mag}$. This estimate is similar to that of P03 using {\em COBE}/DIRBE dust emission data, resulting in an increase in the \ha\ intensity of about $40 \%$. However, because the comparison of \ion{H}{1} column densities and extinction is uncertain and the line ratios of \ha, \nii, and \sii\ are not affected by extinction due to the similar wavelengths of the lines, we do not attempt to apply an extinction correction to our data. In the future, we will obtain \hb\ spectra of the cloud to more accurately measure the \ha\ extinction \citep{mr05}.

\section{Results}

In the \ha, \nii, and \sii\ pointed spectra in Figure~\ref{fig:spectra}, the Smith Cloud is clearly detected near $\vlsr = +100 \kms$. Emission from the warm ionized medium (WIM) in the Saggitarius Arm is evident near $\vlsr = +40 \kms$. In Figure~\ref{fig:smith_map}, we show a map of the \ha\ emission from the Smith Cloud. For each spectrum, we derived intensities from the Gaussian fit described above. The region of brightest \ion{H}{1} emission---the ``tip'' identified by L08---is also brightest in \ha, with the pointed observations indicating an intensity of $0.43 \pm 0.04 \R$. However, a separate component of the cloud near $(l, b) = (38 \arcdeg, -19 \arcdeg)$ (labeled ``clump A'' in Fig.~\ref{fig:smith_clumps}) that is present but faint in \ion{H}{1} has an \ha\ intensity of $0.25 \pm 0.02 \R$.

Along a line from roughly $(38 \arcdeg, -11 \arcdeg)$ to $(43 \arcdeg, -15 \arcdeg)$, the $+100 \kms$ \ion{H}{1} and \ha\ emission both cut off against a narrow ridge; outside this ridge is bright intermediate-velocity \ion{H}{1} emission, identified by yellow contours in Figure~\ref{fig:smith_map}. L08 suggest that this is likely material that has been ram-pressure-stripped from the cloud. At the latitude of the Smith Cloud, the sightlines pass through the Satittarius Arm at $z \approx -1 \kpc$ and $\vlsr \approx +40 \kms$. Due to the $1.0-1.8 \kpc$ scale height of the warm ionized medium \citep[e.~g.][]{hrt99,gmc08,sw09} and the $\lesssim 0.4 \kpc$ scale height of the warm neutral medium \citep{f01}, any \ha\ emission associated with the intermediate-velocity \ion{H}{1} emission cannot be separated from the warm ionized medium \ha\ emission from the foreground Sagittarius Arm at similar velocities.

There is considerable pixel-to-pixel variation in the \ha\ emission on the $1 \arcdeg$ scales probed by WHAM. P03 reported a $20 \%$ variation in \ha\ intensity between two $5'$ beams spaced $0.5 \arcdeg$ apart, and the L08 21~cm data show structure on scales comparable to their $10'$ resolution. Therefore, unresolved \ha\ emission structure may be present in the data presented here.

\subsection{Mass} \label{sec:mass}

\citet{wyw08} estimated the \Hplus\ mass of the Smith Cloud to be $(0.2-1.9) \times 10^6 M_{\odot}$ based on the \ha\ detections in two directions reported by P03. With the cloud mapped in \ha, we refine the mass estimate using one of two assumptions: (1) the \Hplus\ is concentrated in an ionized skin, or (2) the \Hplus\ is fully mixed with the \Hneutral\ gas. For each case, we derive the emission measure from the \ha\ intensity \citep[$\EM \equiv \int n_e^2 \, ds = 2.75 T_4^{0.9} I_{\ha} \pc \cmsix$, where $T_4 = T(10^4 \K)^{-1}$ and $I_{\ha}$ is the \ha\ intensity expressed in Rayleighs;][]{r91}, assuming $T = 10^4 \K$ (see \S~\ref{sec:temp}). If a uniform Lyman continuum flux incident on the cloud maintains the ionization and the cloud is optically thick to Lyman continuum photons, the emission measure is constant across the cloud as seen from the source of the ionizing radiation.

We estimate the mass treating several clumps separately (\S~\ref{sec:skin} and \ref{sec:mixed}) and considering the entire cloud as a contiguous entity (\S~\ref{sec:totalmass}). We first identified each clump, shown in Figure~\ref{fig:smith_clumps}, based on localized \ha\ emission and assigned WHAM beams to the clump by eye. In each case, there is a region of corresponding enhanced \ion{H}{1} emission, so we then chose a minimum \ion{H}{1} column density, $\textrm{Min } \NH$, for each clump which defines the region of enhanced \ion{H}{1}. These values are noted in Table~\ref{tbl:mass}. The ``tip''---which contains the pointed observations discussed above---and ``tail'' are in regions of relatively bright \ion{H}{1} emission, with $\textrm{Min } \NH \sim 10^{20} \cm^{-2}$ and are along the major axis of the cloud. ``Clump A'' and ``clump B'' are in regions with fainter \ion{H}{1}, $\textrm{Min } \NH = 2 \times 10^{19} \cm^{-2}$ and are each separated from the major axis of the cloud by several degrees. These four clumps were chosen subjectively based on their enhancement relative to the diffuse emission. Among these clumps, the tail is most similar in intensity (in both \ha\ and \ion{H}{1}) to the nearby surrounding emission from the cloud.

\subsubsection{Ionized skin} \label{sec:skin}

First, we assume that each clump consists of uniform-density neutral hydrogen with a fully ionized skin of the same temperature and pressure. In this case, the electron density in the skin $n_e = \nneutral / 2$, where \nneutral\ is the density in the neutral clump. This condition is established when the ionized gas within the skin has had time to come to pressure equilibrium with the neutral clump. Alternatively, if this gas has not had time to reach pressure equilibrium, the density in the skin could be as large as that in the neutral clump ($n_e = \nneutral$). We derive physical parameters for an ionized skin model with these two limiting cases below. We further assume that the depth of the neutral cloud along the line of sight is comparable to the projected width, \Lneutral. In the skin, the path length is derived from one of these density assumptions combined with our \ha\ observations: $\Lionized = \EM \, n_e^{-2}$.

The mass of the ionized gas in the cloud within a solid angle $\Omega$ is $1.4 m_\H n_e D^2 \Omega \Lionized$, where $D$ is the distance to the cloud. The mass of a hydrogen atom is $m_\H$; the factor of $1.4$ accounts for helium. The ionized mass in the cloud within one WHAM beam is thus
\begin{equation} \label{eq:mass}
\frac{M_{\Hplus}}{M_{\odot}} = 1.27 \times 10^3 \left(\frac{D}{12.4 \kpc} \right)^2  \frac{\EM}{\pc \cmsix} \left(\frac{n_e}{\mathrm{cm}^{-3}} \right)^{-1}.
\end{equation}

For each clump, we derive the neutral gas density as $\langle n_0 \rangle = \langle \NH \rangle / \Lneutral$, where $\langle \NH \rangle$ is the mean \ion{H}{1} column density over the region where $\NH > \textrm{Min } \NH$. We then use $n_e$ and the observed \EM\ with equation~(\ref{eq:mass}) to calculate the \Hplus\ mass in each beam. The \Hplus\ mass in each clump is then the sum of the masses in the beams identified with the clump. The resulting total \Hplus\ masses and path lengths are listed in Table~\ref{tbl:mass} for both the equal-temperature-and-pressure and the equal-density cases. Using this ionized skin assumption, there is $\approx 1 \times 10^5 M_{\odot}$ of ionized gas in the four clumps. Note that the tip and tail, which contain the brighest \ha\ and 21~cm emission and most of the \Hneutral\ mass, account for $< 20 \%$ of the ionized mass in these clumps: due to the higher density of the gas in the tip and tail, the emission is relatively bright, but \Lionized\ and, therefore, the \Hplus\ mass are relatively small.

The choice of assuming that the neutral and ionized gas are of equal pressure and temperature or equal density results in a factor of two change in the derived mass but a factor of four change in the derived path length of ionized gas. For the tip and tail, the data (Table~\ref{tbl:mass}) are consistent with either equal pressure or equal density: the masses are small and the path lengths much smaller than our projected resolution in both cases. However, for both clumps A and B, equal pressure yields a path length (\Lionized) considerably larger than the projected size of the clump on the sky, whereas equal density yields a value of \Lionized\ similar to the projected size of the clump. Therefore, unless the clumps are elongated along the line of sight by a factor of $2-4$, an ionized skin in pressure equilibrium with a neutral clump does not effectively describe clumps A or B. If the temperature of the neutral gas is lower than the temperature of the ionized gas, the pressure equilibrium ionized skin describes clumps A and B even more poorly. Our data do not rule out an ionized skin of the same density as the neutral gas in clumps A and B. Pressure effects should act in a sound crossing time, $\lesssim 40 \textrm{ Myr}$ for these clumps (assuming $T=10^4 \K$), although external dynamical forces could prevent a static equilibrium from being established in these more tenuous regions \citep{bd97,ppm07,hp09}.

The dominant source of uncertainty within this model is the distance to the cloud. Because $n_e^{-1} \propto \Lneutral \propto D$, the mass scales as $M \propto \EM\, D^3$. The range of allowed distances quoted by L08 yields a $\pm 30 \%$ statistical uncertainty in our mass estimate. Applying an extinction correction would increase the mass. If the $A(\ha) = 0.38 \textrm{ mag}$ we estimated in \S~\ref{sec:extinction} applies throughout the cloud, the mass estimate would be roughly $40 \%$ higher than stated here.

Our derived ionized gas mass estimates are sensitive to the chosen boundaries of the neutral clump and the size of the clump, \Lneutral. Mass estimates scale linearly with \Lneutral, which we choose by eye based on the 21~cm emission. We identify the clumps themselves subjectively, choosing regions with enhanced \ha\ emission and roughly corresponding enhanced 21~cm emission. For the tip, clump A, and clump B, this procedure is reasonably robust; the \ion{H}{1} contours are steep enough so that varying the chosen $\textrm{Min } \NH$ by a factor of $2$ changes the derived \Hneutral\ mass by $\sim 40 \%$. The 21~cm emission defining the tail is considerably less robust, as a number of similar enhancements in \ion{H}{1} are evident in the body of the Smith Cloud (see Fig.~\ref{fig:smith_map}). We identify the cloud based on the strong \ha\ emission and choose the \ion{H}{1} contour level so that the region with the neutral emission roughly matches the region with the ionized emission. However, there is relatively little \Hplus\ mass in the tail, so the chosen contour level is not crucial to the total mass estimate. Alternatively, one could define the \ion{H}{1} boundary of the clump as the extent of the WHAM beams included in the clump. This approach modifies the derived \Hplus\ mass estimates by $\lesssim 30 \%$ in the ionized skin model and $< 10 \%$ in the fully mixed model. Because the \ion{H}{1} data have considerably higher angular resolution than the WHAM \ha\ data, we do not adopt this alternative \ion{H}{1} clump definition.

\subsubsection{Fully mixed ions and neutrals} \label{sec:mixed}

We now estimate the mass of the clumps assuming that the \Hplus\ and \Hneutral\ in each clump are fully mixed or, equivalently, $\Lionized = \Lneutral$. This scenario applies if the clumps consist of a number of smaller, unresolved clouds of uniform density and the source of ionization penetrates or permeates \citep{bsa07} the entire clump. We determine the mean emission measure from the clump, $\langle \EM \rangle$, and derive the root mean square electron density in the clump: $n_e = ( \langle \EM \rangle \Lionized^{-1})^{1/2}$. The ionized mass in the clump is $1.4 m_\H n_e \Lionized \Omega D^2$. The mass of each clump in this model is shown in Table~\ref{tbl:mass}, with a total of $\approx 3 \times 10^5 M_{\odot}$ of ionized gas in the identified clumps. This estimate scales as $M \propto \EM^{1/2} D^{2.5}$, so the uncertainty due to the distance is $\pm 30 \%$ and a $0.38 \textrm{ mag}$ extinction correction would increase the mass by $20 \%$.

\subsubsection{Total mass} \label{sec:totalmass}

\begin{\deluxetablestar}{l r@{$\times 10$}l r@{$\pm$}l r r@{.}l r r@{$.$}l r r@{$.$}l r r@{$.$}l l r@{$.$}l r@{$.$}l}
%\rotate % for aastex only
\tablecolumns{21}
\tablecaption{Mass estimates for individual clumps}
\tablewidth{0pt}
\tablehead{\multicolumn{5}{c}{} & \colhead{WHAM} & \multicolumn{5}{c}{} & \multicolumn{3}{c}{Skin ($n_e = \nneutral / 2$)} & \multicolumn{3}{c}{Skin ($n_e = \nneutral$)} & \colhead{} & \multicolumn{4}{c}{Mixed ($\Lionized=\Lneutral$)} \\
\cline{12-17}
\cline{19-22}
\colhead{Clump} & \mc{Min $\NH$\tnma} & \mc{$\langle \EM \rangle$} & \colhead{beams} & \mc{$M_{\Hneutral}$\tnma} & \colhead{$\Lneutral$} & \mc{$\langle \nneutral \rangle$} & \colhead{$\Lionized$} & \mc{$M_{\Hplus}$\tnmb} & \colhead{$\Lionized$} & \mc{$M_{\Hplus}$\tnmb} & \colhead{} & \mc{$\langle n_e \rangle$} & \mc{$M_{\Hplus}$\tnmc} \\
\colhead{} & \mc{($\textrm{cm}^{-2}$)} & \mc{$(\textrm{pc cm}^{-6}$)} & \colhead{} & \mc{($10^6 M_{\odot}$)} & \colhead{(pc)} & \mc{$(\textrm{cm}^{-3})$} & \colhead{(pc)} & \mc{($10^6 M_{\odot}$)} & \colhead{(pc)} & \mc{($10^6 M_{\odot}$)} & \colhead{} & \mc{$(\textrm{cm}^{-3})$} & \mc{($10^6 M_{\odot}$)} }
\startdata
    Tip&200&$^{18}$ & 0.88 & 0.05 &    3 & 0&6  &  200 & 0&4  &   18 & 0&015 &    5 & 0&008 && 0&07  & 0&05  \\
   Tail& 90&$^{18}$ & 0.31 & 0.07 &    6 & 0&2  &  100 & 0&4  &    8 & 0&012 &    2 & 0&006 && 0&06  & 0&04  \\
Clump A& 20&$^{18}$ & 0.51 & 0.15 &    3 & 0&07 &  200 & 0&06 &  500 & 0&06  &  140 & 0&03  && 0&05  & 0&04  \\
Clump B& 20&$^{18}$ & 0.39 & 0.08 &    6 & 0&19 &  300 & 0&04 &  900 & 0&14  &  200 & 0&07  && 0&04  & 0&08  \\
\hline
Diffuse\tnmd&\mc{\nodata}& 0.11 & 0.01 &  244 & 5&   & 1000 & 0&01 & 4000 & 7&    & 1100 & 3&    && 0&009 & 3&    
\enddata
\label{tbl:mass}
\tablenotetext{a}{\ion{H}{1} column densities are integrated from $+70$ to $+145 \kms$.}
\tablenotetext{b}{Mass estimates in the ionized skin model scale as \EM\ and, therefore, linearly with any extinction correction (\S~\ref{sec:skin}).}
\tablenotetext{c}{Mass estimates in the fully mixed model scale as $\EM^{1/2}$ and, therefore, as the square root of any extinction correction (\S~\ref{sec:mixed}).}
\tablenotetext{d}{``Diffuse'' refers to sightlines not included in the tip, tail, clump A, or clump B.}
\end{\deluxetablestar}

With both the fully mixed model and the ionized skin model, we find that the two clumps with low \ion{H}{1} column density (clumps A and B) each have \Hplus\ masses of $3-8 \times 10^4 M_{\odot}$. In the two clumps with higher \ion{H}{1} column density (the tip and tail), the ionized skin model places much less mass in the tip and tail (each $0.6-0.8 \times 10^4 M_{\odot}$), whereas we find $M_{\Hplus} \approx 4-8 \times 10^4 M_{\odot}$ in each of the four clumps in the fully mixed model.

The clumps identified in our mass estimate do not account for all the \ha\ or \ion{H}{1} emission in the cloud. Using the techniques described above, we calculated the mass in each of the beams not associated with a clump assuming (1) $\nneutral = 0.01 \cucm$, with \Lionized\ derived from the \ha\ and $n_e = \nneutral$ or $n_e = \nneutral / 2$, as in the ionized skin model, and (2) $\Lionized = 1 \kpc$, deriving $n_e$ from the \ha\ emission measure, as in the fully mixed model. We estimated these values based upon the extent of the \ion{H}{1} emission and typical \ion{H}{1} column densities in the regions outside our identified clumps. In the fully mixed model ($\Lionized = \Lneutral$) and equal density ionized skin ($n_e = n_0$) models, the total ionized mass in the diffuse sightlines is $\approx 3 \times 10^6 M_\odot$, considerably more than that present in the clumps. If the temperature and pressure of the ionized and neutral gas are equal, the resulting mass is a factor of $2$ higher. However, we consider this model implausible for this case because the path length along the line of sight, $\Lionized = 4 \kpc$, is much larger than the observed transverse size of the Smith Cloud. These results are all sumarized in Table~\ref{tbl:mass}.

\subsection{Temperature} \label{sec:temp}

Line widths of \ha\ and the relatively narrow \sii\ line allow us to constrain the temperature of the ionized gas, assuming the ionized hydrogen and sulfur are fully mixed \citep{r85b}:
\begin{equation}
\frac{T}{10^4 \K} = \left(\frac{W_\H}{21.1\kms}\right)^2 - \left(\frac{W_\Sulfur}{21.0\kms}\right)^2 - 0.072,
\end{equation}
where $W_\H$ and $W_\Sulfur$ are the full widths at half maximum of the \ha\ and \sii\ lines. This solution requires a distribution of nonthermal velocities with the most probable value
\begin{equation}
\frac{v}{\kms} = \sqrt{\left(\frac{W_\Sulfur}{1.64 \kms}\right)^2 - \left(\frac{W_\H}{9.88 \kms}\right)^2 }.
\end{equation}
The observed widths (Table~1) yield $T = 15500 \pm 7400 \K$ and $v=11.5 \pm 3.4 \kms$ in the direction of the long exposures in the tip. The large uncertainty is primarily due to blending of the the Smith Cloud and Sagittarius Arm emission. If the Sagittarius emission is better physically represented by multiple components, our three-Gaussian fit overestimates $W_\H$ in the Smith Cloud. With a smaller $W_\H$ alone, the temperature of the cloud would be lower. The red wing of the Smith Cloud \ha\ spectrum has nearly zero contribution from the Sagittarius Arm and therefore provides a useful check on the width of the Smith Cloud component. Keeping the \ha\ mean fixed to the $+96.2 \kms$ \sii\ mean, line widths $W_\H < 27 \kms$ fail to reproduce this portion of the spectrum. This yields a lower bound of $T = 7000 \pm 2000 \K$, with the uncertainty propogated from the uncertainty in $W_\Sulfur$.

\subsection{Nitrogen abundance} \label{sec:abundance}

\begin{figure}
\plotone{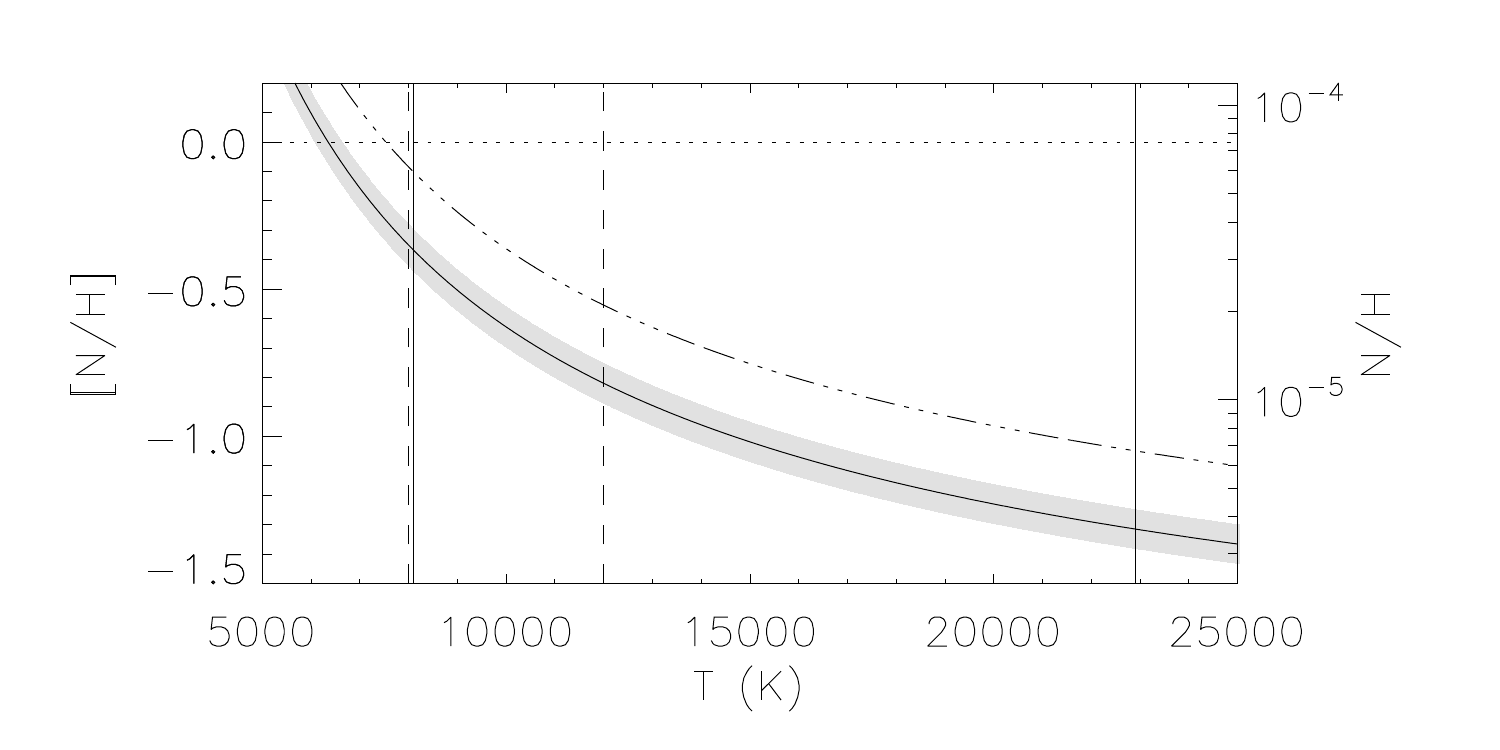}
\caption{Nitrogen abundance ($[\N/\H] \equiv \log(\N/\H) - \log(\N/\H)_{\odot}$) derived from the observed line ratio, $\nii/\ha = 0.32 \pm 0.05$, as a function of the electron temperature, assuming $\N^+/\N = \Hplus/\H$. The error envelope due to the uncertainty in the line ratio is shaded grey. The dot-dashed curve shows the nitrogen abundance derived from the P03 observation of $\nii/\ha = 0.6$. Solid lines show the range of temperatures allowed by the line width analysis, while dashed lines show the range of typical temperatures in a diffuse gas heated by photoionization. The dotted line shows the solar abundance.}
\label{fig:n_abundance}
\end{figure}

We now use the pointed \nii\ and \ha\ observations to constrain the metallicity of the Smith Cloud. The line ratio $\nii \lambda 6583 / \ha$ is sensitive primarily to the temperature and nitrogen abundance of the gas \citep{hrt99}:
\begin{equation} \label{eq:niiha}
\frac{\nii}{\ha} = 1.63 \times 10^5 \left(\frac{\Hplus}{\H}\right)^{-1} \left(  \frac{\N^+}{\N} \frac{\N}{\H} \right) T_4^{0.426} e^{-2.18/T_4}.
\end{equation}
Because of the similar first ionization potentials of nitrogen (14.5~eV) and hydrogen (13.6~eV), $\N^+ / \N \approx \Hplus / \H$; photoionization modelling also supports this argument \citep{shr00}. The observed \oiii\ upper limit of $0.07 \R$ (ionization potential of 35~eV) supports an assumption that there is little $\N^{++}$ (ionization potential of 54~eV) in this gas. Therefore, the observed line ratio $\nii/\ha = 0.32 \pm 0.05$ constrains the nitrogen abundance, shown in Figure~\ref{fig:n_abundance}. The lowest allowed temperature, $\approx 8000 \K$, yields a nitrogen abundance of $\N / \H = 0.4$ times solar (assuming  $(\N / \H)_{\odot} = 7.5 \times 10^{-5}$; \citealt{mcs97}), whereas the highest allowed temperature, $\approx 23000 \K$, yields $\N / \H = 0.05$ times solar. The figure also shows the range of typical temperatures in the WIM, which better constrain the nitrogen abundance with assumptions discussed below.

\section{Discussion and future work} \label{sec:discussion}

Typical \ha\ intensities for HVCs are $0.06-0.5$~R \citep{trh98,hrt01,wvw01,wvw02,pbv03,h05}; the intensities observed here for the Smith Cloud are in this range. The narrow range of \ha\ intensities for these clouds suggests that, like the WIM, all of these HVCs are ionized by a diffuse UV radiation field, not shocks or individual, local hot stars \citep[see also][]{bvc98}, which would produce significant variations in \ha\ intensity from cloud to cloud and even within a cloud. The similar line ratios $\nii/\ha = 0.32 \pm 0.05$ and $\sii/\ha = 0.32 \pm 0.05$ observed here are common in WIM gas but atypical of classical \ion{H}{2} regions \citep{mrh06}.

The extent of the \ha\ emission from the Smith Cloud generally traces the \ion{H}{1}, but we find moderately bright \ha\ emission in regions with faint \ion{H}{1}. This result is similar to that for the intermediate velocity clouds Complex~K and Complex~L, which have also been mapped in \ha\ with WHAM \citep{hrt01,h05}. In all three clouds, the {\em presence} of \ha\ and \ion{H}{1} emission tend to track each other in corresponding sightlines, but there is no strong correlation between the \ha\ and \ion{H}{1} {\em intensities}, with the exception of the tip of the Smith Cloud (blue `$\times$' in Fig.~\ref{fig:smith_map}), which is the brightest portion in both \ha\ and \ion{H}{1}. Note that the peak Smith Cloud \ion{H}{1} column density of $5.2 \times 10^{20} \cm^{-2}$ (L08) is a factor of $10$ higher than that for Complex~K, $5.3 \times 10^{19} \cm^{-2}$ \citep{hrt01}, or Complex~L, $3.6 \times 10^{19} \cm^{-2}$ (P03). The lack of a strong correlation between \ha\ intensity and \ion{H}{1} column density in the low-column density region of the cloud supports previous suggestions that the ionized gas is not mixed with the neutral gas and that variations in the \ha\ intensity outside the high-\ion{H}{1} column density tip may be due to variations in the strength of the ionizing radiation field rather than the distribution of neutral gas. The \ha\ maps of Complex~K, Complex~L, and the Smith Cloud all suggest that ionized gas extends beyond the \ion{H}{1} associated with the clouds. Moreover, there is some evidence for high velocity gas with higher ions, including \ion{C}{4}, along a sightline with no evidence of lower ions or neutral gas \citep{ssl99}.

Our analysis of the \ha\ and \sii\ line widths leaves the temperature of the Smith Cloud very uncertain (\S~\ref{sec:temp}). We estimate constraints on the temperature by comparing the physical conditions in the HVC and the WIM. Typical observed temperatures of WIM gas are $\approx 6000-10000 \K$, with higher temperatures in lower density portions of the WIM \citep{hrt99,mrh06}. Models show that photoionization heating from diffuse UV radiation balanced by radiative cooling does not account for the observed temperatures, suggesting that an additional heating source is important at densities $\lesssim 0.1 \cucm$ \citep{rht99}. Also, a low metallicity would reduce the dominant cooling mechanism, collisionally excited line emission \citep{of06}; the WIM has roughly solar metallicity. The low density and low metallicity of the Smith Cloud suggest that the temperature of the ionized gas in the cloud is $\gtrsim 8000 \K$, the temperature of relatively warm WIM gas. The implied nitrogen abundance is then $0.44 \pm 0.07$ or $0.15 \pm 0.02$ times solar assuming $T = 8000$ or $12000 \K$, respectively (see \S~\ref{sec:abundance} and Fig.~\ref{fig:n_abundance}). For comparison, the metallicity of Complex~C is $0.1-0.2$ times solar \citep[and references therein]{fsw04}.

P03 found a line ratio \nii~$\lambda 6583 / \ha = 0.60$ in each of their fields, significantly higher than the value we find in the $1 \arcdeg$ WHAM beam in the tip of the cloud. We speculate on three possible explanations for this difference: (1) A discrepancy in the data taking, reduction, or analysis, (2) effects of the different beam diameters, $1 \arcdeg$ in our WHAM data, compared to $5'$ in their data, or (3) variations in the nitrogen abundance across the cloud, as the P03 data were taken approximately $2 \arcdeg$ down the major axis of the cloud from the tip, where our pointed spectra were obtained. Beam effects are possible if there is clumping on angular scales not resolved by WHAM. With the typical assumption that individual HVCs are coherent objects with uniform metallicities, a variation in the line ratio along the axis of the cloud would mean that the tip of the cloud is cooler than the downstream portion observed by P03.

Our mass calculations (\S~\ref{sec:mass}) in combination with a comparison of the \ha\ and \ion{H}{1} morphology of the cloud suggest that a different picture is appropriate in the area of the cloud along the major axis (including the tip and tail) than in the area of fainter emission (including clumps A and B). Along the major axis, the data are consistent with an \Hplus\ skin on an optically thick (to ionizing photons) \Hneutral\ cloud: the derived skin depth is small ($\sim 10 \pc$) and the \ha\ emission does not extend considerably beyond the \ion{H}{1}. Meanwhile, in clumps A and B and assuming that the \Hplus\ thermal pressure equals the \Hneutral\ pressure, the required thickness of the ionized layer is considerably larger than the extent of the observed \ha\ emission, implying that the ionized and neutral gas are mixed. This picture indicates that there is relatively little ionized gas associated with the brightest \ha\ and 21~cm emission. The total \Hplus\ mass in the region we have observed is $\sim 3 \times 10^6 M_{\odot}$ (Table~\ref{tbl:mass}). The distributions of the \Hneutral\ and \Hplus\ mass are different: over $80 \%$ of the \Hneutral\ is along the major axis of the cloud, above the green line shown in Figures~\ref{fig:smith_map} and \ref{fig:smith_clumps}, while the \Hplus\ masses above and below the green line are approximaely equal. Because \ha\ emission continues to the edge of our map, we cannot say how far the \ha\ emission (and \Hplus) extends beyond the 21~cm emission.

The kinematics and cometary morphology indicate that the Smith Cloud is moving towards the Milky Way. The immediate question is whether the cloud traces new infall or gas cycling back to the Galaxy as part of a ``Galactic fountain''. L08 argued that, given the trajectory and velocity of the cloud, it is unlikely that the material was ejected from the disk, although the low inclination and prograde orbit of the cloud suggest that the orbit has been affected by the Galactic potential. Our metallicity constraint supports a non-fountain origin.

WHAM began operations from Cerro Tololo Inter-American Observatory in Chile in April 2009. As the Smith Cloud is visible from this southern site as well, we will obtain follow-up observations of the region to expand upon the results presented here. Specifically, we plan to: (1) Extend the \ha\ map beyond the limits presented here. \ha\ emission from the Smith Cloud extends to the edge of the region we have currently mapped. (2) Obtain \hb\ observations that allow more accurate extinction correction (\S~\ref{sec:extinction}), which will improve the estimate of the ionized mass in the cloud. (3) Improve the signal-to-noise ratio of the \oiii\ observations to obtain more stringent constraints on the ionization conditions within the cloud. (4) Add additional, deep observations of \ha, \nii, \sii, and \oiii\ at multiple points in the cloud to improve our understanding of the kinematics of the ionized component and the physical conditions throughout the cloud. Tracing variations in these values should provide considerable insight into the interaction of infalling material with the disk of the Galaxy. In particular, multiple observations of the relatively narrow \sii\ line may allow us to identify variations in the velocity of the ionized gas; we could not identify any variations in the velocity of the Smith Cloud emission with the \ha\ maps presented here (\S~\ref{sec:maps}).

\acknowledgments

WHAM is supported by the National Science Foundation through grants AST~0204973 and AST~0607512. The authors thank F.\ J.\ Lockman for providing the \ion{H}{1} data presented in L08,  R.\ A.\ Benjamin and F.\ J.\ Lockman for helpful discussions, and the anonymous referee for suggestions which led to an improved discussion section.

{\it Facility:} \facility{WHAM}

\bibliography{references}

\notetoeditor{Please place Table~\ref{tbl:fits} and Figure~\ref{fig:spectra} on the same page.}

\end{document}